\documentclass[print]{elsarticle}
\usepackage{lineno,hyperref}
\modulolinenumbers[5]
\usepackage{textcomp}
\usepackage{graphicx}
\usepackage{color}
\journal{Journal of \LaTeX\ Templates}

\begin{document}

\begin{frontmatter}

\title{Measurements of the neutral particle spectra on Mars by MSL/RAD from 2015-11-15 to 2016-01-15}

\author[CAU]{Jingnan Guo\corref{correspondingauthor}}
\cortext[correspondingauthor]{Corresponding author}
\ead{guo@physik.uni-kiel.de}

\author[Leidos]{Cary Zeitlin}
\author[CAU]{Robert Wimmer-Schweingruber}
\author[swri,paris]{Donald M. Hassler}
\author[jan]{Jan K\"ohler}
\author[swri]{Bent Ehresmann}
\author[CAU]{Stephan B\"ottcher}
\author[CAU]{Eckart B\"ohm}
\author[jpl]{David E. Brinza}

\address[CAU]{Christian-Albrechts-Universit\"at zu Kiel, Kiel, Germany}
\address[Leidos]{Leidos, NASA Johnson Space Center, Houston, Texas, USA}
\address[swri]{Southwest Research Institute, Space Science and Engineering Division, Boulder, CO, USA}
\address[paris]{Institut d’Astrophysique Spatiale, Orsay, France}
\address[jan]{Thales Electronic Systems GmbH, Kiel, Germany}
\address[jpl]{Jet Propulsion Laboratory, California Institute of Technology, Pasadena, CA, USA}

\begin{abstract}
The Radiation Assessment Detector (RAD), onboard the Mars Science Laboratory (MSL) rover Curiosity, has been measuring the energetic charged and neutral particles and the radiation dose rate on the surface of Mars since the landing of the rover in August 2012. 
In contrast to charged particles, neutral particles (neutrons and $\gamma$-rays) are measured indirectly: the energy deposition spectra produced by neutral particles are complex convolutions of the incident particle spectra with the detector response functions. An inversion technique has been developed and applied to jointly unfold the deposited energy spectra measured in two scintillators of different types (CsI for high $\gamma$ detection efficiency, and plastic for neutrons) to obtain the neutron and $\gamma$-ray spectra.
This result is important for determining the biological impact of the Martian surface radiation contributed by neutrons, which interact with materials differently from the charged particles. 
These first in-situ measurements on Mars provide (1) an important reference for assessing the radiation-associated health risks for future manned missions to the red planet and (2) an experimental input for validating the particle transport codes used to model the radiation environments within spacecraft or on the surface of planets. 
Here we present neutral particle spectra as well as the corresponding dose and dose equivalent rates derived from RAD measurement during a period (November 15, 2015 to January 15, 2016) for which the surface particle spectra have been simulated via different transport models.    \end{abstract}


\end{frontmatter}


\section{Motivation}
The assessment of the radiation environment on and near the surface of Mars is necessary and fundamental for (1) understanding the risks of radiation-induced biological damage in order to ensure the safety of future manned missions to Mars \citep{cucinotta2011updates} and (2) evaluating the impact of radiation on the preservation of organic bio-signatures \citep[e.g.,][]{dartnell2007modelling}. 
Mars lacks a global magnetosphere, but the atmosphere modifies the spectra of energetic particles -- galactic cosmic rays (GCRs) and solar energetic particles (SEPs) -- that arrive at the top of the atmosphere.    
These particles may create secondary particles via nuclear interactions (spallation and fragmentation) with nuclei in the atmosphere; secondaries may further interact while traversing the atmosphere. The end result is a very complex radiation environment on the surface of Mars \citep[e.g.,][]{saganti2002}. 
Particles reaching the Martian surface may also interact in the regolith and, among other outcomes, produce ``albedo'' particles, including neutrons. Details of the neutron spectra provide information about subsurface hydrogen content, as demonstrated by orbital measurements \citep[e.g.,][]{feldman2002global, boynton2004}. 
Unlike charged particles, neutrons do not undergo ionization energy loss and penetrate through matter easily; they are of considerable concern from the perspective of radiation protection, particularly those in the ``fast'' energy range on the order of MeV, where their biological weighting factors are large \citep{icrp103}. 

\section{Introduction to MSL/RAD}
\begin{figure}[ht!]
\centering
\includegraphics[scale=0.4]{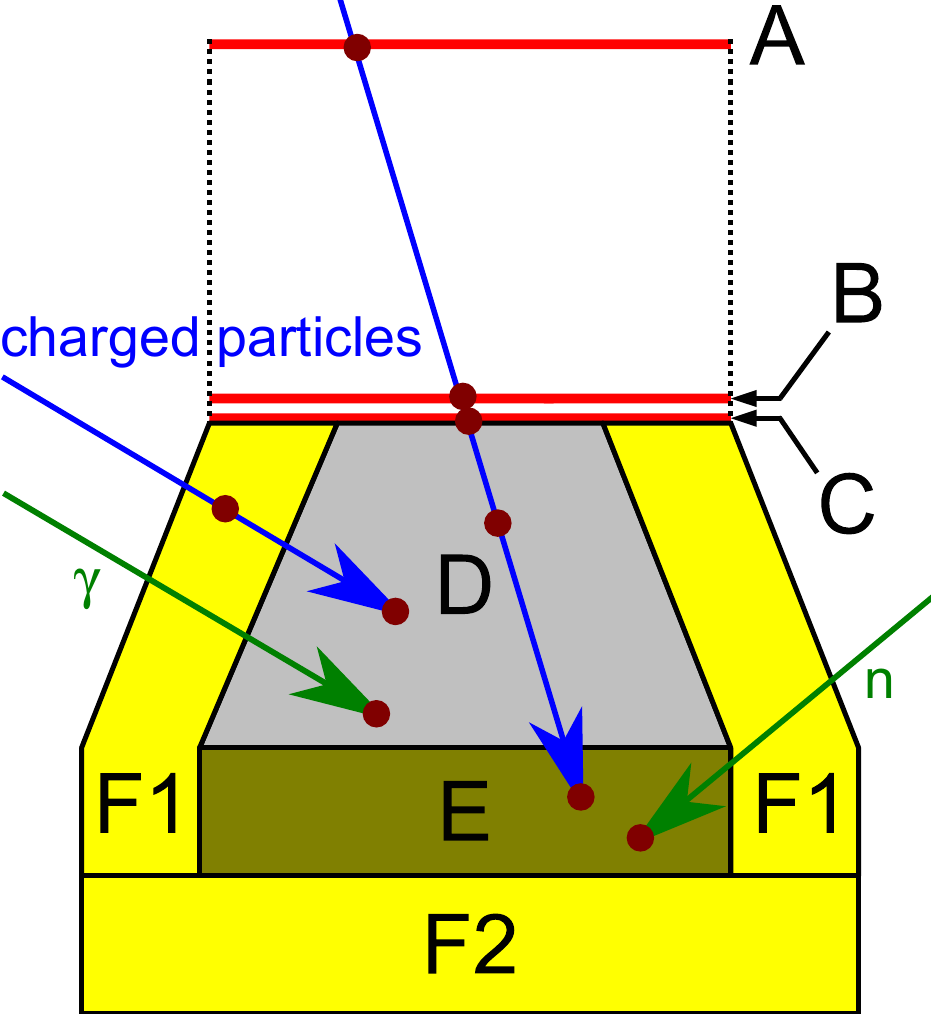}
\caption{Schematic view of the RAD instrument from \citep{guo2015rpd}. RAD consists of three silicon detectors (A, B, C, each having the thickness of 300 \textmu m), a caesium iodide scintillator (D with a height of 28 mm) and a plastic scintillator (E with a height of 18 mm). Both scintillators are surrounded by a plastic anticoincidence (F). For detecting charged particles (blue), A, B, C, D and E are used as a telescope. Neutral particles (both neutrons and gammas, green) are detected in D and E using C and F as anticoincidence.}\label{fig:rad}
\end{figure}

The first-ever measurement on the surface of Mars of both charged \citep{ehresmann2014} and neutral \citep{koehler2014} particles as well as the radiation dose rate \citep{hassler2014} has been carried out by the Radiation Assessment Detector (RAD) \citep{hassler2012} onboard the Mars Science Laboratory (MSL) \citep{grotzinger2012mars} rover Curiosity. 
{A schematic view of the RAD instrument is shown in Fig. \ref{fig:rad} and} a detailed description of the RAD instrument and its scientific objectives can be found in \cite{hassler2012}.
Since the landing of Curiosity on 6 August 2012, RAD has been providing a nearly-continuous set of  measurements from the Martian surface, typically operating in a 16 minute observation cadence. 
The surface radiation has been found to undergo diurnal variations due to the atmospheric thermal tide \citep{rafkin2014}. 
Also, the seasonal CO$_2$ condensation cycle leads to a seasonal\footnote{1 Martian year $\approx$ 668 sols; 1 sol = 1 Mars day $\approx$ 1.03 earth day} variation of the atmospheric column depth above RAD, which also affects the surface radiation \citep{guo2015modeling}.
These are relatively small effects; the solar modulation of primary GCRs has a much stronger effect on the dose rate variations \citep{guo2015modeling}.

RAD contains two scintillators (detector D, made of CsI, and detector E, made of plastic) which are embedded within other plastic scintillators and covered by silicon detectors above. The detectors surrounding D and E are used to define anticoincidence logic, allowing for the measurement of neutral particles in conjunction with charged particle measurements. 
The spectra of charged particles, especially the ones which lose all their energy within the detector sets and are called "stopping particles" (5-100 MeV/nuc for $^1$H and $^4$He), can be measured directly \citep{ehresmann2014}, but the spectra of neutral particles can not be obtained in a straightforward way, since they have a high probability to either pass through the neutral-particle detectors or deposit only a fraction of their energies \citep{koehler2011}. 
Although the high-energy neutrons and $\gamma$-rays that RAD measures are not distinguishable from one another on an event-by-event basis, the energy spectra of the two particle types have been successfully reconstructed on a statistical basis  
\cite{koehler2014}. In that work, the neutral spectra for the first 195 sols (2012-08-06 to 2013-02-21) were obtained using an inversion technique which relies on the modeled detector response function (DRF).  In this study, we carry out a very similar inversion approach and apply it to the more recent data from 2015-11-15 until 2016-01-15.  
In the interim between these measurement periods, several improvements were made to the operating configuration of the instrument: 
\begin{itemize}
\item We now use the onboard neutral histograms based on updated calibration parameters \citep{zeitlin2016calibration} and anticoincidence cuts. The earlier analysis relied on a very small subset of event records (pulse height analyzed events, referred to as PHAs) that were telemetered to Earth. 
\item Use of the onboard histograms means it is no longer necessary to apply a large scale factor, which was not fully-understood at the time of the original analysis. The scale factor is, in essence, the (~ 1\% or less) efficiency for telemetering a neutral-particle event. The onboard histograms provide much better counting statistics compared to the PHA data.
\item The onboard thresholds for the anticoincidence channels F1 and F2 have been carefully adjusted to lie at the center of the noise distributions; this reduces detection efficiency by a well-understood factor of 4 compared to the ideal case, but yields a very clean event sample free from charged-particle contamination.
\item We have accounted for calibration uncertainties in the onboard energy histogram, and we adopt a set of new DRFs which are built with these uncertainties included. 
\end{itemize}  

The neutral-particle triggers require that two (of three) photodiodes used to collect scintillation light record a signal above threshold; to first order, this eliminates background from events in which a $\gamma$-ray deposits some or all of its energy directly in a single photodiode.
Nonetheless, the trigger scheme and onboard analysis are susceptible to overestimating the energy deposited if a $\gamma$-ray deposits energy directly in one of the photodiodes in coincidence with a neutral particle that deposits energy in D and/or E. This affects about 2.4\% of events detected in E and about 0.5\% of events detected in D.
The neutral spectra obtained from data collected in this period are compared with simulated results based on different transport codes (Matthi\"a et al this issue). Because they are all secondary particles, neutral particle spectra allow for detailed comparisons of the radiation transport models that are vitally important for planning future exploration missions. 

\section{The instrument response function}
\begin{figure}[ht!]
\centering
\includegraphics[trim={0 0 23 0}, clip, scale=0.35]{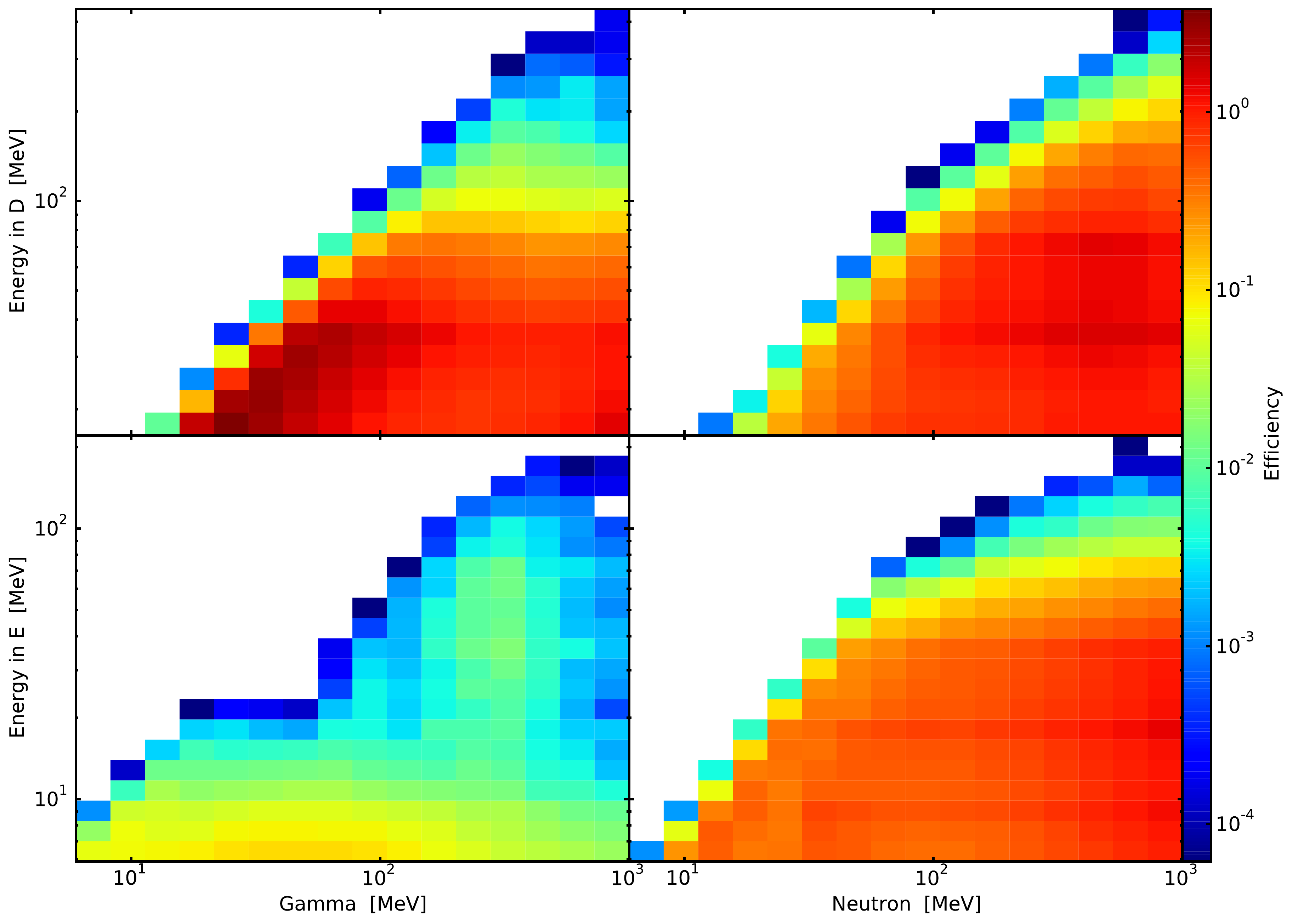}
\caption{The DRF for detecting $\gamma$-rays and neutrons with detectors D and E. The x-axes show the energy of the incoming particles, the y axes show the measured energy in D (top) and E (bottom). The color denotes the detector response (cm$^2$ MeV) for an incoming particle with a given energy in a certain measurement channel. 
The DRF shown here contains 42x32 bins, i.e., 16 bins each for the $\gamma$-ray and neutron spectra, and 42 bins for the D and E measurement data. }\label{fig:drf}
\end{figure}

{As shown in Fig. \ref{fig:rad},} RAD consists of three silicon detectors (A, B, C), a cesium iodide scintillator (D), and a plastic scintillator (E) and both scintillators are surrounded by a plastic anticoincidence (F). 
For detecting downward-going charged particles, A, B, C, D, and E are used as a telescope. 
Neutral particles including $\gamma$-rays and neutrons are detected in D and E using the surrounding detectors C and F to veto any charged particles entering the scintillators. Our anticoincidence logic (AC) is defined as a hit in D and/or E with no hits above threshold in C or F.
When a neutral particle is detected by the scintillators, the deposited energy cannot be easily related to the energy of the incident particle, but rather can range from zero up to the full energy of the particle. That is, the neutral particle spectra in D and E do not directly reflect the incident $\gamma$-ray and neutron spectra. 

The detector response function (DRF) represents the probability of the deposited energy distribution (measured histogram) to be produced by a neutral particle with a particular incident energy. The DRF has been constructed by \cite{koehler2011, koehler2014} using GEANT4 simulations \citep{agostinelli2003} and is shown in Fig. \ref{fig:drf}. 
The simulation is based on a detailed model of instrument and includes effects such as electronic and optical noise in D, E, and F.
{A total of 10$^8$ particles (gammas or neutrons) with isotropically distributed incident directions were used in the GEANT4 simulation and the energy deposit in different RAD detectors was recorded for each incident particle. In reality, the neutron flux seen by RAD is likely anisotropic (which could also be energy-dependent) since (1) the Martian surface upward albedo spectra are different from the downward spectra as they are generated differently in the soil \citep{deangelis2007} and (2) the upward flux is further modified as going through the rover body. Different directionality of the incident flux based on model predictions could be potentially employed for the simulations and the constructions of the DRF. 
However we would like to minimize the influence of models on our measurement-based inversion process, we do not pursue further of the dependence of the DRF on the field directionality.} 

{The unit of the DRF comes from the GEANT4 simulation setups. Specifically, a number of particles within certain energy range (x-axes in Fig. \ref{fig:drf}) is fed into the simulation and the output detected energy histogram has the unit of counts cm$^2$ MeV which accounts for not just the production probability but also the detection area and the chosen energy range of the simulation input. We normalize the histogram by dividing it by the number of input particles so that it has the unit of counts/particle cm$^2$ MeV as shown in y-axes of the DRF. }

The energy range of the incident spectra (x axes) was 6-1000 MeV for both $\gamma$-rays and neutrons. Minimum energies were selected so that a significant fraction of $\gamma$-rays and neutrons would create energy deposits above the detection thresholds. Lower-energy neutral particles escape detection by RAD.
A maximum energy of 1000 MeV is selected; this is somewhat arbitrary, but the expected neutral particle spectra decreases rapidly at such high energy (since the spectrum roughly follows a power-law shape), and detection efficiency falls with increasing energy (due to the increasing probability that a recoil proton or Compton electron escapes D or E and deposits significant energy in C or F).
Of course, particles with energies above this range (e.g., a 2 GeV neutron) can create, e.g., a 50 MeV energy deposit in E without causing a hit in the AC. 
This is not simulated/included in our DRF, but is still expected from measurement although with much lower (a few orders of magnitude) probabilities.
The energy range of the deposited energy histogram (y axes in Fig. \ref{fig:drf})  was 16-442 MeV for D and 5-221 MeV for E. The lower limit of the energy range is restricted in order to avoid the contamination of the spectra by $\gamma$-rays produced by Curiosity's radioisotope thermoelectric generator. 
The upper limit is above the maximum energy deposit found in the calibration and simulations. 

This DRF is a combined matrix \textbf{A} with 4 sub-matrices: $A_{D,\gamma}$ (upper left of the figure), $A_{E,\gamma}$ (lower left), $A_{D,n}$ (upper right), and $A_{E,n}$ (lower right). 
Assuming the spectra of the incoming neutral particles are $f_\gamma$ and $f_n$ for $\gamma$-rays and neutrons respectively, the expected measured histogram in D would be a combination of histograms induced by both $\gamma$-rays and neutrons as :
\begin{eqnarray}\label{eq:zD=Axf}
\vec{z}_D = A_{D,n} \cdot f_n +  A_{D,\gamma} \cdot f_\gamma.
\end{eqnarray}
Similarly, the expected measured histogram in E is :
\begin{eqnarray}\label{eq:zE=Axf}
\vec{z}_E = A_{E,n} \cdot f_n +  A_{E,\gamma} \cdot f_\gamma.
\end{eqnarray}
Directly solving the above equations is challenging due to the ill-posed nature of the inversion problem. A DRF matrix in general may be noninvertible, and in any case inversion yields non-unique results that tend to have large, negative bin-to-bin correlations. A minimization method is preferable, and can be applied with the following physically-reasonable constraints: (1) both $f_\gamma$ and $f_n$ larger than zero in each bin; (2) the flux ratio for adjacent energy bins must not be larger than 10 or less than 0.1. 
The minimization function based on a Gaussian statistics of measurement is :
\begin{eqnarray}\label{eq:chi2_constrain}
{\rm{min}} \sum_{i}^{} (\frac{A \cdot f - z_i}{\sigma_i^2}), 
\end{eqnarray}
where $z_i$ is the measurement in each bin and $\sigma_i$ is the standard error of the measurement. 

\section{Measured neutral histograms}
\begin{figure}[ht!]
\includegraphics[trim={20 10 40 20}, clip, scale=0.40]
{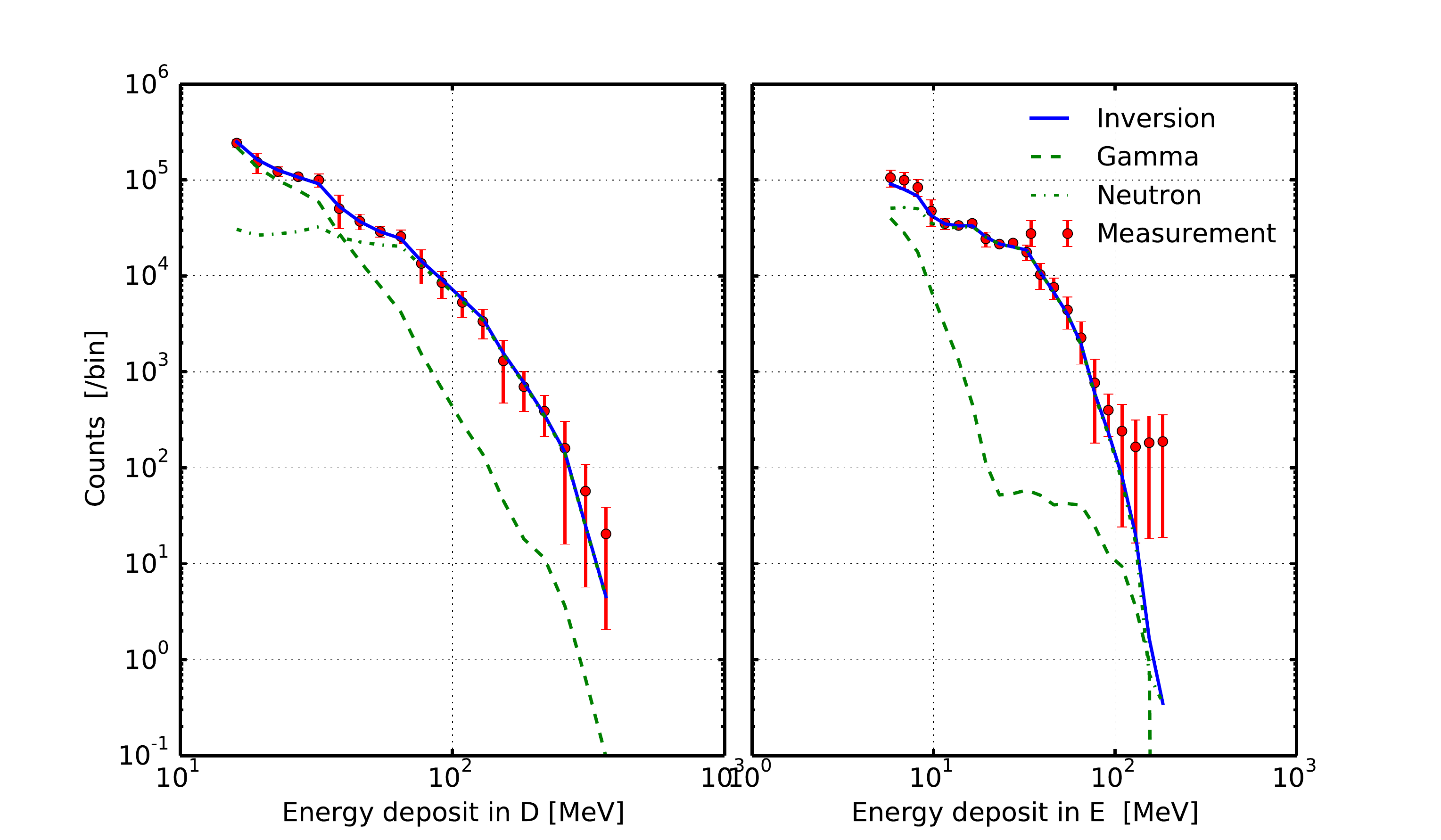}
\caption{Neutral particle measurements from the Martian surface (red) in the scintillators D (left) and E (right) and the result $A \cdot f$ (blue) by multiplying the inverted spectra with the DRF. The individual contribution by $\gamma$-rays ($A_{D/E,\gamma} \cdot f_\gamma$) and neutrons ($A_{D/E,n} \cdot f_n$) are shown in green dahsed and dash-dotted lines respectively.  }\label{fig:histogram}
\end{figure}

Previously in the work of \cite{koehler2014}, the PHA data were used to construct histograms of energy deposited in D and E. Each telemetered event was assigned a weight according to its storage priority, a value which is assigned in the onboard event processing, and which ranges from 0 to 3. 
Neutral particle events were (and continue to be) assigned priority 0 in almost all cases. Only a very small subset of priority-0 PHA events is sent back to Earth, whereas nearly 100\% of higher-priority events are telemetered. The weighting factor for each priority bin was determined by the ratio of the number of events in that bin (as recorded in the onboard software) to the number actually telemetered. Subsequent to publication of \cite{koehler2014}, it was discovered that the count kept onboard included a category of events that should have been excluded, and that our priority-0 weighting factor was overestimated by roughly a factor of two. 
{The events that should have been excluded are "no-readout" events in which a fast trigger is caused by a direct $\gamma$-ray hit in a photodiode, but there is no energy deposit in the scintillators. We have now implemented a counter which keeps track of such events.}

In the current study, we use the onboard stored and telemetered neutral-particle histograms. 
Events triggered in D and/or E without creating signals above threshold in the AC are registered as neutral events and stored in the neutral histogram. These histograms are stored once per observation, which is typically 16 minutes in duration.  
The neutral histograms of both D and E measured from 2015-11-15 to 2016-01-15 over 60 sols (4582683 seconds of integrated observation time) are shown in Fig. \ref{fig:histogram} in red. 
The error bars shown include not only the counting statistics but also the uncertainties (larger in comparison) of the energy scales in D and E.  
The non-linearity of light output from the scintillators D and E has been discussed in detail in \cite{zeitlin2016calibration}, where new calibration parameters of the detectors suggest that the energy measured in both scintillators has an uncertainty of about 10\%. 
The calibration uncertainty cannot be accounted for simply by adding a 10\% error bar in each bin of the onboard histogram; rather we 'reshuffled' the histogram along the energy axis, giving each count a 10\% uncertainty in its energy. This yields conservative estimates of the errors in each bin.

The error bars of the last three bins of each D and E histogram were further enlarged to account for another issue that arises in the inversion process. This issue, referred to as the overflow problem, is caused by particles with incident energies larger than 1 GeV which were not considered in the simulations used to build the DRF. Such particles, although expected to have much lower fluxes than the simulated particles, still contribute to our measured histograms, especially in the highest deposited energy bins. 
As they are excluded from the DRF as well from the inverted spectra, their contamination of the measured histogram should also be minimized, and we force this by reducing the weights of the last three bins (with large $\sigma_i$) \footnote{The uncertainties of the last three bins were chosen to be 0.9 of the counts in the corresponding bin. Such a value 0.9 is pure empirical as the inverted spectra stay more or less unaffected as we further enlarge $\sigma_i$.} in the minimization process shown in Eq. \ref{eq:chi2_constrain}.

\section{Inverted $\gamma$-ray and neutron spectra}
\begin{figure}[ht!]
\includegraphics[trim={0 0 0 20}, clip, scale=0.40]
{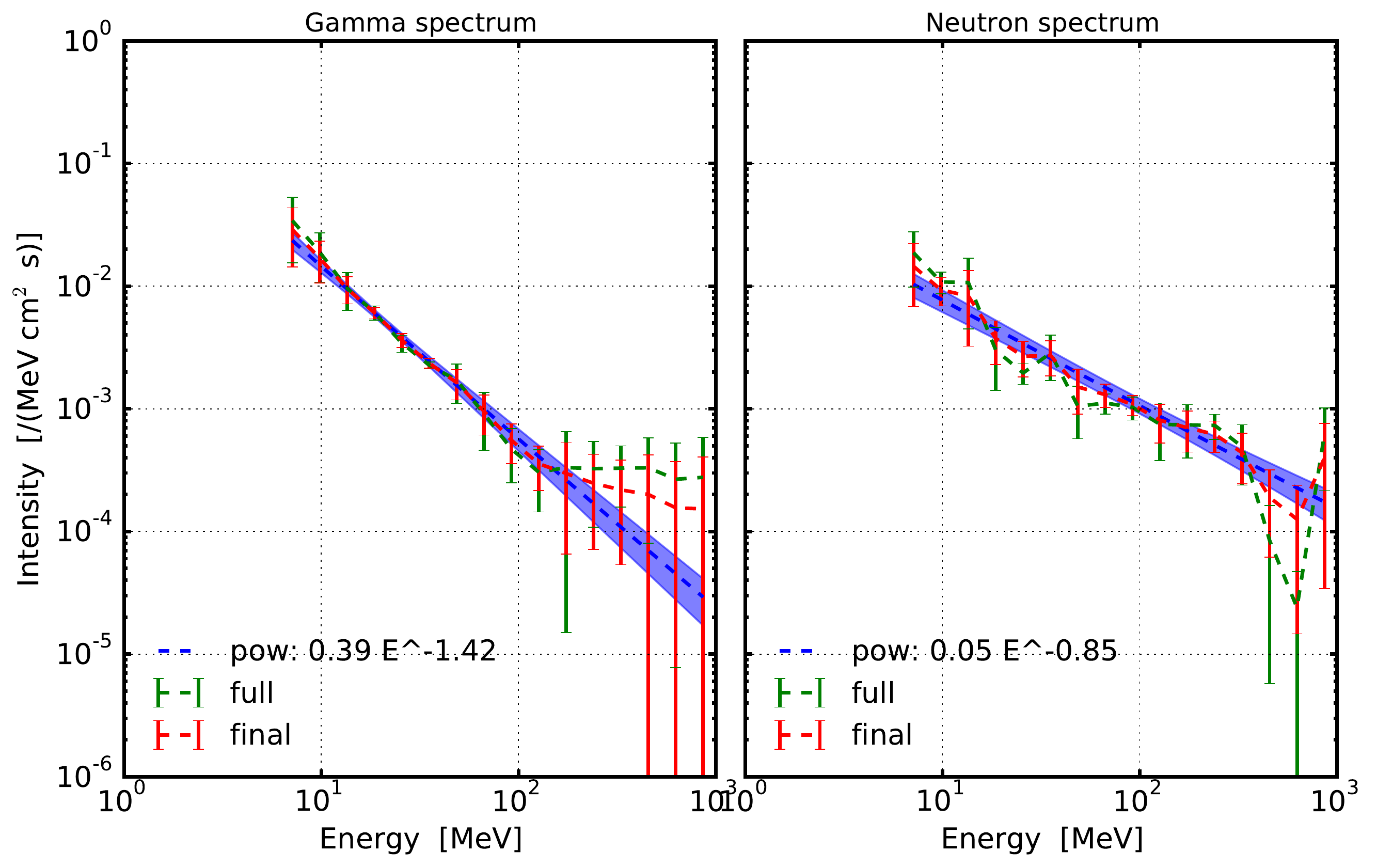}
\caption{$\gamma$-ray (left) and neutron (right) spectra from power law inversion (blue) and full inversion (green) as well as the final combined inversion (red).
The shaded area denotes the error of the power law inversion. }\label{fig:spectra}
\end{figure}

Based on the DRF matrix \textbf{A}, the measured histogram $\vec{z}$, and its error, we can minimize the function in Eq. \ref{eq:chi2_constrain} with given physical constrains to obtain the optimized particle spectra $f_\gamma$ and $f_n$. 
We have found that the inverted results are very sensitive to (a) the initial guesses and (b) the uncertainty in the measured histogram. 

To address point (a), we use a large set of initial guesses to ensure that the inversion yields a result close to the global minimum rather than a local minimum. The result with the smallest error value is then used as the most likely result \citep{koehler2011}.  
Initial guesses of both the neutron and $\gamma$-ray spectra have power law shapes with spectral indices in a range of 1.1-1.9 for $\gamma$-rays and 0.5-1.2 for neutrons \citep{koehler2014}.

To address point (b), in addition to using the ``reshuffled'' histogram, we have also reconstructed several new DRFs in such a way that the deposited energy in each bin is also re-distributed with a 10\% random error. Each of the DRF is also used for the inversion. 
The final error of the inverted spectra was determined via a bootstrap Monte Carlo approach \citep{press2007numerical} where results from different initial guesses and different DRFs were accounted for.  

\subsection{Power-law inversion}
In the relevant energy range, the gamma and neutron spectra can be approximated by power laws as indicated by modeled results \citep[e.g.,][]{deangelis2007} with the following form:
\begin{eqnarray}\label{eq:power-law}
f_{n/\gamma} = I_{n/\gamma} \cdot E^{S_{n/\gamma}}, 
\end{eqnarray}
where $I_{n/\gamma}$ is the spectral intensity for neutral/$\gamma$-ray spectra scaled at 1 MeV and $S_{n/\gamma}$ is the spectral index. 
The inversion of a two-parameter power-law is much faster than a full inversion in which each bin of the spectra $f_i$ is a free parameter during the optimization process. 

Figure \ref{fig:spectra} shows the $\gamma$-ray (left) and neutron (right) spectra from power law inversion in blue. The shaded range marks the error range of the results based on boot-strap Monte Carlo simulations with statistic and calibration error as well as different initial guesses all accounted for. 
The averaged power-law spectral indices are:\\
\begin{tabular}{cc}
$I_\gamma=$ 0.39 $\pm$ 0.07 /(cm$^2$ s MeV) & $S_{\gamma}=$ -1.42 $\pm$ 0.05\\ 
$I_n=$ 0.05 $\pm$ 0.01 /(cm$^2$ s MeV)  & $S_{n} =$ -0.85 $\pm$ 0.03.
\end{tabular}

The slopes of the above spectra agree very well with previous results by \cite{koehler2014} where $S_{n} = 0.72 \pm 0.07$ and $S_{\gamma} = 1.43 \pm 0.1$. 
However the spectral intensities for both types of particles are smaller than in the previous results. That is, the newly obtained spectra have similar shapes to those found previously, but are significantly lower in their intensities. Both the solar modulation and the atmospheric pressure conditions are very comparable for the two measurement periods. 
The analysis based on histograms constructed from the PHA data (as carried out in \cite{koehler2014}) was repeated for this period, and results are in good agreement with results obtained using the onboard histograms.
Therefore the discrepancy is almost certainly due to the priority-0 weighting problem described above. 

\subsection{Full inversion}
We have also carried out the full inversion considering the flux in each bin $f_i$ as an unknown parameter when optimizing Eq .\ref{eq:chi2_constrain}. 
Again, different DRFs were considered and for each DRF a large set of the initial guesses based on power law spectra was tried. 
Figure \ref{fig:spectra} also shows full inversion results of the $\gamma$-ray (left) and neutron (right) spectra in green.
As explained above, the measurements also contain particles from energies above 1 GeV which appear mainly in the last bins as overflow bins and this is visible in both the $\gamma$-ray and neutron spectra (only last bin). 
In other energy ranges, the power-law inverted spectra and fully inverted spectra agree reasonably well, especially for the $\gamma$-ray spectra. 
A small bump is visible in the neutron spectra at $\sim 15$ MeV, as well as a dip at $\sim 500$ MeV; these features are also visible in modeling results \citep[e.g,][]{deangelis2007}. 
Detailed discussions of the inverted spectra as compared with models will be presented in Matthi\"a et al (this issue). 

In considering the accuracy of the results, it is important to realize that representing the spectra with power laws may be a fairly crude approximation; on the other hand, the bin-by-bin inverted spectra is more affected by the overflow issue. 
Therefore we combined the results of these two methods into the final spectra as shown in Fig. \ref{fig:spectra} in red. The error bars include all the uncertainties based on boot strap Monte Carlo simulations and propagated through the analysis. 

\section{Neutron dose and dose equivalent rate}
\begin{table}[htb!]
\centering
\caption{Dose rate and dose equivalent rate of the inverted Martian neutron spectrum (7-740 MeV) from 2015-11-15 to 2016-01-15}\label{table:dose}
\begin{tabular}{c|cc}
& Dose rate & Dose equivalent rate \\
Power law inversion & 5.6 $\pm$ 0.8 \textmu Gy/d & 25.3 $\pm$ 3.3 \textmu Sv/d\\
Full inversion   & 4.7 $\pm$ 0.9 \textmu Gy/d & 22.0 $\pm$ 4.1 \textmu Sv/d\\
Averaged final  & 5.1 $\pm$ 1.0 \textmu Gy/d & 23.6 $\pm$ 4.1 \textmu Sv/d\\
Mean total dose measured  & 233 $\pm$ 12 \textmu Gy/d & 610 $\pm$ 45 \textmu Sv/d\\
\hline
\end{tabular}
\end{table}
Based on the final inverted neutron spectrum, we calculated the dose and dose equivalent of particles within this energy range (7-740 MeV) using United States Nuclear Regulatory Commission [2009] {(https://www.nrc.gov/reading-rm/doc-collections/cfr/part020/full-text.html)} {where the values of the quality factor and the conversion (fluence per unit dose equivalent) factor are calibrated using a 30-cm diameter cylinder tissue-equivalent phantom. The procedure is as follows: apply the above conversion factor to the neutron fluence bin-by-bin to get the dose equivalent histogram; each dose equivalent value is then divided by the energy-dependent quality factor to obtain a dose histogram; the histograms are finally integrated to get the total dose and dose equivalent.}
The results are shown in Table \ref{table:dose}, where the averaged total dose rate measured by RAD during this time period is also shown. 
Dose and dose equivalent rates derived from power law inversion, full bin-wise inversion, and the final averaged are shown in the first three rows. 

It shows that the dose rate result obtained with power law inversion is, within error bars, mostly consistent with but slightly higher than that from full inversion. This is probably due to the dips of the full inverted spectra between $\sim$ 17 and $\sim$ $50$ MeV, a range where the neutron radiaiton weighting factor is high. 
The last row of the table shows the averaged dose and dose equivalent rates measured by the plastic scintillator E during this period.  
The 7-740 MeV neutron contributed dose rate is about 2.2\% of the total dose rate; neutron contributed dose equivalent rate is about 3.8\% of the total dose equivalent rate. 
These values are smaller compared to those obtained by \citep{koehler2014},  likely due to the newly updated histogram \citep{zeitlin2016calibration} as discussed earlier.

A simple independent check of neutral-particle dose rates has been obtained directly from the neutral-particle histograms. Treating D and E separately, the onboard deposited energy histograms were used to determine the total energy deposition in each detector during the measurement period. These were then converted from units of MeV to Joules and divided by the total data acquisition time and appropriate mass to get the uncorrected dose rate in each detector. These results were multiplied by a factor of 4 to account for the AC efficiency described above (trigger thresholds at the centers of the noise distributions). The D neutral dose rate obtained by this method was 7.5 $\mu$Gy/day, and the E neutral dose rate was 5.9 $\mu$Gy/day; {there may be small contributions to both of these rates from the radioisotope thermo-electric generator (RTG), but they are assumed to be negligible since the large majority of the RTG $\gamma$-ray flux is below the detection thresholds which are $\sim$ 7 MeV in D and $\sim$ 3.5 MeV in E. } Since this simple method inherently incorporates both neutron and $\gamma$-ray doses, and given the high efficiency of D for detecting $\gamma$-rays, it is not surprising that the D dose rate is larger than that in E. Based on the cumulative energy deposited distributions and spectra presented in \cite{koehler2014}, we estimate that roughly 25\% of the dose rate in D is due to neutrons (in particular those with E $\ge$ 50 MeV), and that about 90\% of the dose in E is due to neutrons. Adding the two results gives 6.7 $\mu$Gy/day, though we expect this is a slight overestimate due to double-counting. Even so, the estimate is in line with the neutron dose rate found by the inversion methods above.

\section{Discussion and Conclusion}
We have used an inversion method following \cite{koehler2014} to determine the Martian gamma and neutron spectra from the MSL/RAD neutral particle measurements from 2015-11-15 to 2016-01-15. 
The gamma and neutron spectra were obtained by two variations on the inversion method, one using power law spectra and the other a bin-wise full inversion. 
Within the estimated errors, the results obtained by the two inversion methods agree reasonably well with each other.
The spectral shapes found here agree very well with previous results \citep{koehler2014}, but the intensities of both the neutron/$\gamma$-ray spectra are smaller compared to earlier results obtained during very similar solar modulation and atmospheric conditions. The disparity is attributable to the improved methodologies employed in this analysis {especially the usage of the onboard histograms which have been better calibrated without priority-0 weighting problems}.

The neutron dose rate and the dose equivalent over the inverted energy range were calculated for the two inversion methods and compared to the average plastic dose rate measured by RAD on the surface of Mars during this period. 
The 7-740 MeV neutron contributed dose rate is only about 2.2\% of the total dose rate while the corresponding dose equivalent rate is about 3.8\% of the total dose equivalent rate. 
Based on the power law results, we extrapolate our inverted neutron spectra to 1-1000 MeV range and calculated the dose and dose equivalent to be 6.76 \textmu Gy/day and 31.24 \textmu Sv/day, corresponding to $\sim$ 3\% of the total dose rate and $\sim$ 5\% of the charged-particle dose equivalent rate.

Since neutrons and $\gamma$-rays on the Martian surface are all secondary particles, the neutral spectra inverted here are unique experimental data for verifying the particle transport models which are vitally important for planning future exploration missions. This comparison is carried out by Matthi\"a et al in this issue.

\section*{Appendix}
In order to facilitate the usage of the neutral spectra presented in Fig. \ref{fig:spectra}, we have listed the values of the spectra in Table \ref{table:spec}. 

\begin{table}[htb!]
\centering
\caption{Inverted $\gamma$ and neutron spectra as well as their uncertainties shown in Fig. \ref{fig:spectra} (red). }\label{table:spec}
\begin{tabular}{c|cc|cc|}
\hline
Energy & $f_\gamma$ & $\sigma f_\gamma$  & $f_n$ & $\sigma f_n$\\
MeV & /(MeV cm$^2$ s) & /(MeV cm$^2$ s)& /(MeV cm$^2$ s)& /(MeV cm$^2$ s) \\
\hline
7.13 & 2.90$\times 10^{-2}$ & 1.75$\times 10^{-2}$ & 1.45$\times 10^{-2}$ & 1.01$\times 10^{-2}$\\ 
9.82 & 1.70$\times 10^{-2}$ & 9.39$\times 10^{-3}$ & 9.33$\times 10^{-3}$ & 3.43$\times 10^{-3}$ \\ 
13.52 & 9.56$\times 10^{-3}$ & 3.69$\times 10^{-3}$ & 8.32$\times 10^{-3}$ & 5.19$\times 10^{-3}$ \\
18.61 & 6.06$\times 10^{-3}$ & 7.27$\times 10^{-4}$ & 3.75$\times 10^{-3}$ & 1.48$\times 10^{-3}$ \\
25.62 & 3.63$\times 10^{-3}$ & 5.05$\times 10^{-4}$ & 2.68$\times 10^{-3}$ & 8.82$\times 10^{-4}$ \\
35.27 & 2.36$\times 10^{-3}$ & 2.50$\times 10^{-4}$ & 2.72$\times 10^{-3}$ & 8.97$\times 10^{-4}$\\
48.56 & 1.64$\times 10^{-3}$ & 4.75$\times 10^{-4}$ & 1.51$\times 10^{-3}$ & 6.31$\times 10^{-4}$\\
66.86 &  9.54$\times 10^{-4}$ & 3.67$\times 10^{-4}$ & 1.30$\times 10^{-3}$ & 3.33$\times 10^{-4}$\\
92.05 & 5.55$\times 10^{-4}$ & 2.17$\times 10^{-4}$ & 1.08$\times 10^{-3}$ & 2.62$\times 10^{-4}$ \\
126.74 &  3.57$\times 10^{-4}$ & 1.59$\times 10^{-4}$ & 8.07$\times 10^{-4}$ & 3.14$\times 10^{-4}$\\
174.49 &  2.97$\times 10^{-4}$ & 2.43$\times 10^{-4}$ & 7.01$\times 10^{-4}$ & 2.78$\times 10^{-4}$\\
240.23 &  2.47$\times 10^{-4}$ & 1.89$\times 10^{-4}$ & 6.19$\times 10^{-4}$ & 1.98$\times 10^{-4}$\\
330.75 &  2.19$\times 10^{-4}$ & 1.79$\times 10^{-4}$ & 4.39$\times 10^{-4}$ & 2.04$\times 10^{-4}$\\
455.37 &  2.00$\times 10^{-4}$ & 2.35$\times 10^{-4}$ & 1.90$\times 10^{-4}$ & 1.36$\times 10^{-4}$\\
626.94 &  1.56$\times 10^{-4}$ & 2.33$\times 10^{-4}$ & 1.26$\times 10^{-4}$ & 1.13$\times 10^{-4}$\\
863.17 & 1.53$\times 10^{-4}$ & 2.78$\times 10^{-4}$ & 3.96$\times 10^{-4}$ & 3.73$\times 10^{-4}$\\
\hline
\end{tabular}
\end{table}

\section*{Acknowledgments}
RAD is supported by the National Aeronautics and Space Administration (NASA, HEOMD) under Jet Propulsion Laboratory (JPL) subcontract \#1273039 to Southwest Research Institute and in Germany by DLR and DLR's Space Administration grant numbers 50QM0501 and 50QM1201 to the Christian Albrechts University, Kiel. 
Part of this research was carried out at JPL, California Institute of Technology, under a contract with NASA. 
The data used in this paper are archived in the NASA Planetary Data System’s Planetary Plasma Interactions Node at the University of California, Los Angeles. The archival volume includes the full binary raw data files, detailed descriptions of the structures therein, and higher-level data products in
human-readable form. The PPI node is hosted at http://ppi.pds.nasa.gov/.

\section*{References}

\bibliography{msl_rad_guo}

\end{document}